\documentclass[pra,aps,floatfix,amsmath,twocolumn]{revtex4}

\usepackage{amssymb}
\usepackage{graphicx}
\usepackage{graphics}
\usepackage{amsmath}
\usepackage{amsthm}
\usepackage{color}
\usepackage{dsfont}
\usepackage{CJK}
\usepackage{mathtools}
\def\ra{\rangle}
\def\la{\langle}

\newtheorem{theorem}{Theorem}

\newtheorem{pro}[theorem]{Proposition}

\newcommand{\bea}{\begin{eqnarray}}
\newcommand{\eea}{\end{eqnarray}}
\newcommand{\be}{\begin{equation}}
\newcommand{\ee}{\end{equation}}
\newcommand{\ba}{\begin{equation}\begin{aligned}}
\newcommand{\ea}{\end{aligned}\end{equation}}

\theoremstyle{remark}

\def\be{\begin{equation}}
\def\ee{\end{equation}}

\newcommand{\mH}{\mathcal{H}}

\newcommand{\mS}{\mathcal{S}}

\newcommand{\lr}{\rangle\langle}

\newcommand{\tr}{{\rm Tr}}

\newcommand{\mbb}[1]{\mathbb{#1}}

\def\>{\rangle}
\def\<{\langle}

\begin{document}
	\preprint{APS/123-QED}

	\title{Strict entanglement monotonicity under local operations and classical communication}

	\author{Yu Guo}
	\email{guoyu3@aliyun.com}
	\affiliation{Institute of Quantum Information Science, School of Mathematics and Statistics, Shanxi Datong University, Datong, Shanxi 037009, China}

	\begin{abstract}
	Entanglement monotone is defined as a convex measure of entanglement that
	does not increase on average under local operations and classical communication (LOCC). 
	Here we call an entanglement monotone a \textit{strict entanglement monotone} (SEM) if it decreases strictly on average under LOCC.
	We show that, for any convex roof extended entanglement monotone that on
	pure states is given by a function of the reduced states, if the function is strictly concave, then it is a SEM.
	Moreover, we prove that the negativity and the relative entropy of entanglement
	which are not defined by the convex roof structure, are also SEMs. 
	In addition, if the squashed entanglement could be obtained by some optimal extension, then it is a SEM as well.
	Our results imply that entanglement is strictly decreasing on average under LOCC.

	\end{abstract}
	
	\maketitle

	
Entanglement is one of the most crucial features
of quantum theory as compared to classical theory,
which is also considered to be a valuable resource for quantum information
processing~\cite{Nielsenbook,Wildebook}. 
To quantify the amount of entanglement contained in 
a composite quantum system
is a
fundamental problem in quantum information science and
quantum physics~\cite{Horodecki2009,Guhne,Plenio2007,Donald2002jmp}.
The first significant milestone in this field came from
the discovery that entanglement 
can be used as a resource for distributed quantum information processing in
the frame work of
local operations and classical communication (LOCC)~\cite{Bennett1996}. 
Consequently, to identify certain \textit{a priori} axioms for a good measure of 
entanglement, Vedral \textit{et al.}~\cite{Vedral1997} proposed three conditions for a quantity to be such a measure for the first time.
Later, Vidal in Ref.~\cite{Vidal2000} explored a 
more restrictive requirement on LOCC, and an additional demand of convexity is needed, and there the satisfactory
measure is called an entanglement monotone. 

It is interesting that these constraints on entanglement measures
can be easily checked~\cite{Vidal2000}: For any convex roof extended entanglement measure, 
it is an entanglement monotone if it can be defined by both a locally unitary invariant 
and a concave function on the reduced states of the pure states [see Eqs.~\eqref{h} and~\eqref{concave} below]. 
Recently, we found that, for almost all entanglement measures so far, 
the associated functions are not only concave, but also strictly concave~\cite{GG2}. 
More significantly, this strict concavity guarantees the monogamy of entanglement~\cite{GG2} 
where the monogamy law is a key feature of entanglement distribution 
among multiparties (see Refs.~\cite{GG2,GG,Dhar} and references therein for details).
This motivates us to investigate entanglement measures deeply.
In this paper, we investigate this strict concavity in a more general sense: We show that entanglement
is strictly monotonic under LOCC on average for many entanglement monotones.
That is, we exploit here a new property of the entanglement monotone.

Let $\mH^{A}\otimes\mH^B\equiv\mH^{AB}$ be a bipartite Hilbert space with finite dimension, where $A,B$ are subsystems of the composite quantum system, 
and let $\mS(\mH^{AB})\equiv\mS^{AB}$ be the set of density operators acting 
on $\mH^{AB}$.
Recall that, a function $E: \mS^{AB}\to\mbb{R}_{+}$ is called a 
measure of entanglement if it satisfies~\cite{Vedral1997}: (E1) $E(\rho)=0$ iff $\rho$ is separable [this condition can be replaced by $E(\rho)=0$ if $\rho$ is separable];
(E2) $E$ is invariant under local unitary operations, i.e.,
$E(\rho)=E(U_A\otimes U_B\rho U_A^\dag\otimes U_B^\dag)$ for any local
unitaries $U_A$ and $U_B$;
(E3) $E$ cannot increase under LOCC, i.e.,
$E[\Phi(\rho)]\leq E(\rho)$ for any LOCC $\Phi$.
Note that (E3) implies (E2).
The map $\Phi$ is completely positive and trace preserving (CPTP). 
In general, LOCC can be stochastic in the sense that $\rho$ can be converted to $\sigma_{j}$ with some probability $p_j$. In this case, the map from $\rho$ to $\sigma_{j}$ can not be described in general by a CPTP map. 
More than
(E2), $E$ is said to be an entanglement monotone~\cite{Vidal2000} if it is  
nonincreased on average under stochastic LOCC, i.e.,
\be\label{average}
E(\rho)\geq\sum_jp_jE\left( \sigma_j\right) ,\quad \forall\,\rho\in\mS^{AB}.
\ee
Note that Eq.~\eqref{average} is
more restrictive than $E(\rho)\geq E(\sum_ip_j\sigma_j)$
since
in such a case we cannot select subensembles according to a measurement
outcome~\cite{Plenio2005}. 	
It is possible that $E(\sigma_{j_0})>E(\rho)$ for some $j_0$.
Almost all measures of entanglement studied in literature satisfy~\eqref{average}.
The measure is said to be faithful if it is zero only on separable states.

	Let $E$ be a measure of entanglement on bipartite states.  
	The entanglement of formation $E_F$ associated  with $E$ is defined by
	\begin{align}
	E_F\left(\rho\right)\equiv\min\sum_{j=1}^{n}p_jE\left(|\psi_j\lr\psi_j|\right),\label{eofmin}
	\end{align}
	where the minimum is taken over all pure state decompositions of $\rho=\sum_{j=1}^{n}p_j|\psi_j\lr\psi_j|$.
	That is, $E_F$ is the convex roof extension  of $E$.
	Vidal~\cite[Theorem 2]{Vidal2000} showed that $E_F$ above is an entanglement monotone on mixed bipartite states  
	if the following concavity condition holds. For a pure state $|\psi\rangle\in\mH^{AB}$, $\rho^A={\rm
		Tr}_B|\psi\rangle\langle\psi|$, define the function $h:\mS^{A}\rightarrow\mathbb{R}_+$ by
	\be\label{h}
	h\left( \rho^A\right) \equiv E\left( |\psi\lr\psi|\right).
	\ee 
	Note that since $E$ is invariant under local unitaries we must have
	$h\left( U\rho^AU^\dag\right) =h\left( \rho^A\right)$
	for any unitary operator $U$ acting on $\mH^A$. If $h$ is also concave, i.e.
	\be \label{concave}
	h[\lambda\rho_1+(1-\lambda)\rho_2]\geq\lambda h(\rho_1)+(1-\lambda)h(\rho_2)
	\ee
	for any states $\rho_1$, $\rho_2$, and any
	$0\leq\lambda\leq1$, then $E_F$ as defined in~\eqref{eofmin} is an entanglement monotone.
	
	It was shown in Ref.~\cite{GG2} that for almost all the well-known entanglement measures, the associated function 
	defined as in~\eqref{h} is strictly concave (from which we proved that 
	$E$ is monogamous on pure tripartite states and $E_F$ is 
	monogamous on both pure and mixed tripartite states, according to our definition in Ref.~\cite{GG}). Then, in the sense of Vidal~\cite{Vidal2000},
	what is the corresponding property of LOCC if $h$ is strict concave?
	We introduce here the concept of \textit{strict entanglement monotone}
	in terms of more restriction on the LOCC in (E3). We then show that $E_F$ is a strict
	entanglement monotone if the associated function $h$ is strict concave.
	Going further, we will prove that many entanglement measures, such as the {negativity}~\cite{VidalWerner}, the relative entropy of entanglement~\cite{Vedral1997,Vedral1998pra}, and the squashed entanglement~\cite{Christandl2004jmp} 
	(if it can be obtained by some optimal extension)
	are strict entanglement monotones. 
	Our results would demonstrate that entanglement measures are strict in our sense.

	For convenience, we fix some terminologies. 
An entanglement measure $E$ is said to be strictly decreasing on average
under LOCC if
for any stochastic LOCC, 
\begin{eqnarray}
&&\left\lbrace \Phi_j:{\rm Tr}\Phi_j(\rho)=p_j,~\sum_jp_j=1, ~\right.\quad\quad\quad\quad\nonumber \\
&&\quad\quad\quad\Phi_j(\rho)\neq p_jU_j^A\otimes U_j^B\rho {(U_j^A)}^\dag\otimes {(U_j^B)}^\dag, \quad\nonumber\\
 &&\quad\quad\quad\left. ~~~{(U_j^X)}^\dag U_j^X=I_j^X,~ j=1,~ 2,~\dots,~ d.\right\rbrace\quad 
\end{eqnarray} 
there exists $\rho\in\mS^{AB}$ such that 
\be\label{strict1}
E\left( \rho\right)>\sum_jp_jE\left( \sigma_j\right),
\ee
where $p_j\sigma_j=\Phi_j(\rho)$, $U_j^{X}$ are unitary operators on $\mH^{X}$.
Equivalently, an entanglement measure $E$ decreases strictly on average under LOCC if and only if 
\be\label{strict2}
E\left( \rho\right)=\sum_jp_jE\left( \sigma_j\right)
\ee 
holds for all states $\rho\in\mS^{AB}$ implies that
the LOCC is either a local unitary operation
(if the LOCC is a map from system $A+B$ to $A'+B'$, then it is a local isometric operation; hereafter, we always assume with
no loss of generality that the LOCCs are acting from $A+B$ to itself) or a convex mixture of local unitary operations.
If an entanglement monotone $E$ is strictly decreasing on average under LOCC, we call it is a \emph{strict
	entanglement monotone} (SEM).
If an entanglement monotone $E$ is strictly decreasing under LOCC for pure states,
we call it is a SEM on pure states.

	\begin{theorem}\label{p-m}
		Using the notations above, if $E$ is a SEM on pure states, then $E_F$ is a SEM as well.
	\end{theorem}
	
	\begin{proof}
	 According to the LOCC scenario in Ref.~\cite{VidalWerner},
	 in order to prove that a
	 local unitary invariant function $E: {\mathcal S}^{AB}\rightarrow{\mathbb
	 	{R}}_+$
	 satisfying condition (E1) is an entanglement monotone, 
	 we only need to consider a family $\{\Phi_k\}$ consisting of completely positive linear maps
	 such that $\Phi_k(\rho)=p_k{\sigma_k}$, $\rho\in\mS^{AB}$,
	 where $\Phi_k(X)=I^A\otimes M_kXI^A\otimes M_k^\dag$ transforms pure states to some scalar multiple of pure states,
	 $\sum_kM_k^\dag M_k=I^B$.
	 
	 Applying $\Phi_{k}$ to $\rho$, the state becomes
	\be\sigma_k={\Phi_{k}(\rho)}/{p_k}\nonumber
	\ee
	with probability $p_k={\rm Tr}\Phi_{k}(\rho)$. \if false Therefore, the
	final state is $\sigma=\sum_kp_k\sigma_k$. By \cite{Vidal2000}, if
	$E(\sigma)\leq E(\rho)$ holds for $\Phi_{k}$, then $E_F$ is an entanglement monotone (it is clear that $E_F$ is convex). \fi
	We assume that $\rho=|\psi\ra\la\psi|\in\mS^{AB}$ is an entangled pure state.  
	It yields
	\begin{eqnarray}\label{pureineq}
		E(|\psi\ra\la\psi|)\geq\sum_kp_kE\left( \sigma_k\right)=\sum_kp_kE_F\left( \sigma_k\right).
	\end{eqnarray}
	If $E$ is a SEM on pure states and the equality holds in~\eqref{pureineq} for any pure state $|\psi\ra\in\mH^{AB}$,
	then either $\Phi_{k}\equiv\Phi^B$ for some local unitary operation $\Phi^B$ or $\Phi_{k}(\cdot)=p_kI^A\otimes U_k^B(\cdot)I^A\otimes (U_k^B)^\dag$.

	Now we assume that $\rho$ is mixed. Perform $\Phi_{k}$ on $\rho$
	and denote $\sigma_k=\Phi_{k}(\rho)/p_k$ with
	probability $p_k={\rm Tr}\Phi_{k}(\rho)$. Observe that there exists an ensemble
	$\{t_j,|\eta_j\rangle\}$ of $\rho$
	such that
	\begin{eqnarray*}
		E_F(\rho)=\sum_jt_jE\left( |\eta_j\rangle\right) .
	\end{eqnarray*}
	For each $j$, let
		$\sigma_{jk}=\frac{1}{t_{jk}}\Phi_{k}\left( |\eta_j\rangle\langle\eta_j|\right)$,
	where $t_{jk}={\rm
		Tr}\Phi_{k}(|\eta_j\rangle\langle\eta_j|)$. Then
		$\sigma_k=\frac{1}{p_k}\sum_jt_jt_{jk}\sigma_{jk}$
	and
		$E(|\eta_j\rangle)\geq\sum_kt_{jk}E\left( \sigma_{jk}\right)$ 
	by what is proved for pure states above. 
	It follows that
	\begin{eqnarray}
		&&E_F(\rho)=\sum_jt_jE\left( |\eta_j\rangle\right) \nonumber\\
		&\geq&
		\sum_{j,k}t_jt_{jk}E\left( \sigma_{jk}\right) 
		\geq\sum_kp_kE_F\left( \sigma_k\right) .
	\end{eqnarray}
	If $E_F(\rho)=\sum_kp_kE(\sigma_k)$ for any $\rho\in\mS^{AB}$, then $E(|\eta_j\ra)=\sum_kt_{jk}E_F(\sigma_{jk})$,
	which completes the proof by the result of the case for pure states.
	\end{proof}

	\begin{pro}\label{condofSEM}
     $E_F$ as defined in~\eqref{eofmin} is a SEM 
     if the associated function $h$ in Eq.~\eqref{h} is strictly concave,  i.e, 
     $h[\lambda\rho_1+(1-\lambda)\rho_2]>\lambda h(\rho_1)+(1-\lambda)h(\rho_2)$ whenever $\rho_1\neq\rho_2$, $0<\lambda<1$.
	\end{pro}
	
	\begin{proof} 
	We only need to check it for pure states by Theorem~\ref{p-m}.
	We use the notations as in the proof of Theorem~\ref{p-m} and we assume without loss of generality that $k=1$, 2.

	If $h$ is strictly concave, we assume that the equality holds in~\eqref{pureineq}, which leads to
	\begin{eqnarray}\label{pureineq2}
	h(\sigma^A)=h\left( \sum_kp_k\sigma^A_k\right) =\sum_kp_kh\left(\sigma^A_k \right)
	\end{eqnarray}
	since
	$E(|\psi\ra\la\psi|)=h(\rho^A)=h(\sigma^A)$
	and $\sum_kp_kh\left(\sigma^A_k \right)
	= \sum_kp_kE\left( \sigma_k\right)=\sum_kp_kE_F\left( \sigma_k\right)$,
	where $\rho^A={\rm Tr}_B|\psi\ra\la\psi|$,
    $\sigma^A_k={\rm Tr}_B\sigma_k$ and
	$\sigma^A=\sum_kp_k\sigma^A_k$.
	Then $\sigma^A_k=\sigma^A_l$ for any $k$ and $l$, which implies that
	either $\Phi_{k}\equiv\Phi^B$ for some local unitary operation $\Phi^B$ or $\Phi_{k}(\cdot)=p_kI^A\otimes U_k^B(\cdot)I^A\otimes (U_k^B)^\dag$, where $U_k^B$'s are unitary operators on $\mH^B$, $\sum_kp_k=1$.
	The proof is completed.
	\end{proof}

Note that, many entanglement measures, such as entanglement of distillation $E_d$, entanglement cost $E_c$, the squashed entanglement $E_{\rm sq}$~\cite{Christandl2004jmp}, and the relative entropy of entanglement $E_r$ coincide with the entanglement of formation $E_f$ (hereafter, we denote by $E_f$ the original entanglement formation~\cite{HillWotters}) for pure states~\cite{Donald2002jmp,Vedral1998pra,Christandl2004jmp}.
In addition, $E_d\leq E_{\rm sq}\leq E_f$~\cite{Christandl2004jmp}, $E_r\leq E_f$~\cite{Vedral1998pra}, and $E_c\leq E_f$~\cite{Christandl2003,Hayden2001jpa}. 
Thus $E_f$, $E_c$, $E_{\rm sq}$, and $E_r$ are SEMs on pure states,
and $E_d$ decreases strictly under LOCC on average for pure states ($E_d$ is not an entanglement monotone since it is not convex, see Table~\ref{tab:table1}).

\begin{theorem}
	Let $E$ be an entanglement monotone that
	for pure states it is defined as in~\eqref{h}. 
	Then $h$ is strictly concave if and only if for any
	stochastic LOCC $\{\Phi_j: 
	\Phi_j(\cdot)=I^A\otimes M_j(\cdot)I^A\otimes M_j^\dag\}$ and any pure state $\rho\in\mS^{AB}$ that satisfies
	\be\label{monogamycond}
	\sigma_{j_0}^A\neq\rho^A\,
	\ee
	for some $j_0$ we have~\eqref{strict1} holds,
	where $p_j\sigma_j=\Phi_j(\rho)$, $X^A={\rm Tr}_BX$.
\end{theorem}

\begin{proof}
	The `only if' part is clear.
	Conversely, if~\eqref{monogamycond}
	holds, it is equivalent to say that
	if $E(\rho)=\sum_jp_jE(\sigma_j)$ then we must have $\sigma_j^A=\rho^A$ for any $j$.
	Note that $\sigma_j$'s are pure states, it follows that
	for any pure state $\rho\in\mS^{AB}$, $h(\rho^A)=\sum_jp_jh(\sigma_j^A)$ if and only if
	$\sigma_j^A=\rho^A$ for all $j$.
	That is, $h$ is strictly concave.	
\end{proof}

It is interesting that we can give another proof of part 1 in Ref.~\cite[Theorem]{GG2} from condition~\eqref{monogamycond}.
We recall part 1 of the Theorem in Ref.~\cite{GG2}: \textit{Let $E$ be an entanglement monotone for which $h$, as defined in Eq.~\eqref{h}, is strictly concave. If $\rho^{ABC}=|\psi\lr\psi|^{ABC}$ is pure and $E(\rho^{A|BC})=E(\rho^{AB})$, then 
$\mH^B$ has a subspace isomorphic to $\mH^{B_1}\otimes\mH^{B_2}$ and up to local unitary on system $B_1B_2$,
\be\label{product}
|\psi\ra^{ABC}=|\phi\ra^{AB_1}|\eta\ra^{B_2C}\,,
\ee
where $|\phi\ra^{AB_1}\in\mH^{AB_1}$ and $|\eta\ra^{B_2C}\in\mH^{B_2C}$ are pure states.
In particular, $\rho^{AC}$ is a product state [and, consequently, $E(\rho^{AC})=0$], so that $E$ is monogamous on pure tripartite states.}
In order to see this, we 
	let $\{|i_b\ra\}$ and $\{|j_c\ra\}$ be orthonormal bases of	$\mH^B$
	and $\mH^C$, respectively.
	Define
	\be
	V_j|\psi\ra\equiv\sum_i\la i_b|\la j_c|\psi\ra|i_b\ra,\quad \forall~|\psi\ra\in\mH^{BC}.
	\ee
	It follows that
	\be
	{\rm Tr}_C\rho^{ABC}=\sum_j I^A\otimes V_j\rho^{ABC}I^A\otimes V_j^\dag.
	\ee	
	Let $\rho^{ABC}=|\psi\ra\la\psi|^{ABC}$
	and assume that it satisfies $E(\rho^{A|BC})=E(\rho^{AB})$, $\rho^{AB}={\rm Tr}_C\rho^{ABC}$.
	Let $\{|k_a\ra\}$ be an orthonormal basis of $\mH^A$,
	then
	\be\nonumber
	|\psi\ra^{ABC}=\sum_{k,i,j}a_{kij}|k_a\ra|i_b\ra|j_c\ra.
	\ee
	The action of $\Phi_s(\cdot)\equiv I^A\otimes V_s(\cdot)I^A\otimes V_s^\dag$
   on $\rho^{ABC}$ gives
   \bea
   p_s\rho_s^{AB}=\Phi_s\left( \rho^{ABC}\right) =|\psi'_s\ra\la\psi_s'|,
   \eea		
	where
	$|\psi'_s\ra=\sum_{k,i}a_{kis}|k_a\ra|i_b\ra$.
	That is, $\rho_s^{AB}$ is a pure state for any $s$.
    On the other hand, $E$ obeys~\eqref{monogamycond}, which results in
    \be
    \rho_s^A=\rho^A,\quad\forall\,s,
    \ee
	where $\rho_s^A={\rm Tr}_B\rho_s^{AB}$.Note that $\rho^{AB}=\sum_s|\psi_s\ra\la\psi_s|^{AB}$,
	then
	following the proof of the Theorem in Ref.~\cite{GG2},
	we can conclude that  
	$\mH^B$ has a subspace isomorphic to $\mH^{B_1}\otimes\mH^{B_2}$ and up to local unitary on system $B_1B_2$,
	\be\label{product}
	|\psi\ra^{ABC}=|\phi\ra^{AB_1}|\eta\ra^{B_2C}\,,
	\ee
	where $|\phi\ra^{AB_1}\in\mH^{AB_1}$ and $|\eta\ra^{B_2C}\in\mH^{B_2C}$ are pure states.

	In what follows, we discuss whether or not the entanglement monotones that are not derived via the convex roof structure are SEMs as well.
	The well known one is the computable measure of entanglement, negativity, 
	which is defined by~\cite{VidalWerner} 
	\be
	N(\rho)=\dfrac{\left\| \rho ^{T_A}\right\|_{\rm Tr}-1}{2}\,,\quad \rho\in\mS^{AB},
	\ee
	where $\|X\|_{\rm Tr}=\tr\sqrt{X^\dag X}$ and $\rho^{T_A}$ denotes the partial transposition with respect to part $A$ under some given orthonormal bases of $\mH^A$ and $\mH^B$.
	The logarithmic negativity $E_N$ is defined
	as~\cite{VidalWerner}
	\be
	E_N(\rho)=\log_2N(\rho)\,.
	\ee
	It is known that the negativity $N$ is a SEM on pure states~\cite{GG2} and thus
	$N_F$ is also a SEM by Proposition~\ref{condofSEM}.
	In what follows we will show that $N$ is also a SEM on mixed states, and thus it is a SEM.

   \begin{theorem}
   	The negativity $N$ is a SEM. 
   \end{theorem}

	\begin{proof}
		According to the scenario in Ref.~\cite{VidalWerner},
		we only need to consider a family $\{\Phi_k\}$ consisting of completely positive linear maps
		such that $\Phi_k(\rho)=p_k{\sigma_k}$,
		where $\Phi_k(X)=I^A\otimes M_kXI^A\otimes M_k^\dag$ transforms pure states to some scalar multiple of pure states,
		$\sum_kM_k^\dag M_k= I^B$. For any $\rho\in\mS^{AB}$ with $N(\rho)>0$,
		 we let
		 \be
		 {\rho}^{T_A}=(1+a)\rho^+-a\rho^-,
		 \ee
		where $(1+a)\rho^+$ and $a\rho^-$ are the positive part and the negative part of ${\rho}^{T_A}$, respectively.
		That is, $N(\rho)=a$, $\rho^+\rho^-=\rho^-\rho^+=0$.
		It follows that
		\begin{eqnarray}
		p_k\sigma_k^{T_A}&=&\Phi_k(\rho)^{T_A}=\Phi_k\left( \rho^{T_A}\right) \nonumber\\
		&=&(1+a)\Phi_k\left( \rho^+\right) -a\Phi_k\left( \rho^-\right).
		\end{eqnarray} 
		It is clear that $N(\sigma_k)\leq q_ka/p_k$, $q_k={\rm Tr}\Phi_k(\rho^-)$.
		Thus, if Eq.~\eqref{strict2} holds, then $N(\sigma_k)= q_ka/p_k$, and thus
		$\Phi_k(\rho^+)\Phi_k(\rho^-)=\Phi_k(\rho^-)\Phi_k(\rho^+)=0$.
		Take $\rho=|\psi\ra\la\psi|$ with
		$|\psi\ra=\sum_j\lambda_j|j_a\ra|j_b\ra$ as the 
		Schmidt decomposition of $|\psi\ra$.
		Then $|\psi\ra\la\psi|^{T_A}=\sum_j\lambda_j^2|j_a\ra\la j_a|\otimes |j_b\ra\la j_b|+\sum_{i<j}\lambda_i\lambda_j|\psi_{ij}^+\ra\la\psi_{ij}^+|-\sum_{i<j}\lambda_i\lambda_j|\psi_{ij}^-\ra\la\psi_{ij}^-|$,
		where $|\psi_{ij}^{\pm}\ra=\frac{1}{\sqrt{2}}(|i_a\ra|j_b\ra\pm|j_a\ra|i_b\ra)$.
		Denoting by $M_k|j_b\ra\equiv|j'_b\ra$,
		it follows that
		$\la i'_b|i'_b\ra=\la j'_b|j'_b\ra$
		holds for any $i$ and $j$, from which we can conclude that $M_k$ is a scalar multiple of some unitary operator. 
		This guarantees 
		that $\Phi_k$ is either
		 a local unitary operation or $\Phi_{k}(\cdot)=q_kI^A\otimes U_k^B(\cdot)I^A\otimes (U_k^B)^\dag$ with $\sum_kq_k=1$
		provided that $N(\rho)=\sum_kp_kN({\sigma_k})$.
		Therefore, $N$ decreases strictly on average under LOCC.
	\end{proof}

	\begin{pro}	
	  The logarithmic negativity $E_N$ decreases strictly under LOCC on average, but it is not a SEM.
	\end{pro}
   
   \begin{proof}
   It is clear that $E_N$ decreases strictly under stochastic LOCC on average since the logarithm is strictly concave.
   But $E_N$ is not convex~\cite{Plenio2005}, namely, it is not an entanglement monotone, therefore it is not a SEM. 
   \end{proof}

   Another important entanglement monotone that
   is not derived from the convex roof extension
   is the relative entropy of entanglement~\cite{Vedral1997,Vedral1998pra}:
   \be
   E_r(\rho^{AB})\equiv \min\limits_{\sigma^{AB}}S\left( \rho^{AB}||\sigma^{AB}\right) ,
   \ee
   where $S(\rho^{AB}||\sigma^{AB})\equiv{\rm Tr}[\rho^{AB}(\ln\rho^{AB}-\ln\sigma^{AB})]$ is the quantum relative entropy and the minimum is taken over all separable states $\sigma^{AB}$ in $\mS^{AB}$. This
   measure,
   as one might expect, is a SEM.

	\begin{table*}
		\caption{\label{tab:table1} A list of  entanglement measures.
			We denote the distillable entanglement~\cite{Plenio2006qic}, entanglement cost~\cite{Bennett1996pra}, entanglement of formation (the original one defined in Ref.~\cite{HillWotters}), concurrence~\cite{Rungta2003pra}, $G$ concurrence~\cite{Gour2005}, negativity,
			convex roof extended negativity~\cite{Lee}, the logarithmic negativity,
			tangle, squashed entanglement, Tsallis-$q$ entanglement~\cite{Kim2010pra},  
			R\'{e}nyi-$\alpha$ entanglement~\cite{Gour2007jmp,Kimand}, the relative entropy of entanglement and
			the conditional entanglement of mutual information~\cite{Yang2008prl} by $E_d$, $E_c$, $E_f$,
			$C$, $G$, $N$, $N_{F}$, $E_N$, $\tau$, $E_{\rm sq}$, $T_q$, $R_\alpha$,
			$E_r$ and $E_I$, respectively.}	
		\begin{ruledtabular}
			\begin{tabular}{ccccccccc}
				$E$              & Continuity  &Additivity             &Convex      & Faithfull  & Relation  & $h$   & Monogamy  & Strict decreasing\footnotemark[1]\\ \colrule
				$E_d$            &  ?\footnotemark[2]            &$\times$\cite{Shor2001}&$\times$\cite{Shor2001} &$\times$    &$\leq E_c$&Strict concave & All states\footnotemark[3]     & Pure states\\
				$E_c$            &   ?          &  $\checkmark$                     &   $\checkmark$\cite{Donald2002jmp}         &   ?        &$\leq E_f$\cite{Christandl2003,Hayden2001jpa}&Strict concave& Pure states   &Pure states\\
				$E_f$            &$\checkmark$ &?                &$\checkmark$&$\checkmark$&          &Strict concave& All states&$\checkmark$\\
				$C$              &$\checkmark$ &                       &$\checkmark$&$\checkmark$&          &Strict concave& All states&$\checkmark$\\
				$G$              &$\checkmark$ &                       &$\checkmark$&$\checkmark$&          &Strict concave & All states&$\checkmark$\\
				$N$              &$\checkmark$ &                       &$\checkmark$&$\times$    &          &Strict concave& Pure states     &$\checkmark$\\
				$N_F$            &$\checkmark$ &                       &$\checkmark$&$\checkmark$&          & Strict concave& All states&$\checkmark$ \\
				$E_N$         &$\checkmark$ &                       &$\times$    &$\checkmark$&          & Strict concave& All states&$\checkmark$ \\
				$\tau$           &$\checkmark$ &                       &$\checkmark$&$\checkmark$&          & Strict concave& All states&$\checkmark$\\
				$E_{sq}$\cite{Christandl2004jmp}&$\checkmark$\cite{Alicki} &$\checkmark$           &$\checkmark$&$\checkmark$&$E_d\leq 
				E_{sq}\leq
				E_c$       &Strict concave&All states&Pure states\footnotemark[4]\\
				$T_{q}$, $ q>0$    &$\checkmark$ &                       &$\checkmark$&$\checkmark$&          &Strict concave& All states&$\checkmark$\\
				$R_{\alpha}$, 
				$0 \leq\alpha\leq1$&$\checkmark$ &                       &$\checkmark$&$\checkmark$&          &Strict concave& All states&$\checkmark$\\
				$E_r$              &$\checkmark$ &$\times$\cite{Vollbrecht}               &$\checkmark$&$\checkmark$&$\leq E_f$\cite{Vedral1998pra}&Strict concave& Pure states     &$\checkmark$\\
				$E_I$
				~\cite{Yang2008prl}&$\checkmark$ &$\checkmark$           &$\checkmark$&?     &$\leq E_f$,$E_c$&Strict concave\footnotemark[5]&Pure states      &Pure states
			\end{tabular}
		\end{ruledtabular}
		\footnotetext[1]{Here strict decreasing refers to the strict decreasing property of the measure under LOCC on average.}
		\footnotetext[2]{? means it is unknown.}
		\footnotetext[3]{The one-way distillable entanglement is monogamous~\cite{Koashi}}
		\footnotetext[4]{For mixed states, see Theorem~\ref{7}.}
		\footnotetext[5]{It is easy to check that $E_I(|\psi\ra\la\psi|^{AB})=S(\rho^A)$ for any pure state $|\psi\ra^{AB}$, $\rho^A={\rm Tr}_B|\psi\ra\la\psi|^{AB}$.}
	\end{table*} 		
		
    \begin{theorem}
    	$E_r$ is a SEM.
    \end{theorem}

    \begin{proof}
    	Let $\mH^C$ be an extended Hilbert space of $\mH^{AB}$, let $\{|i_c\ra\}$ be an orthonormal basis in $\mH^C$, and let $|\alpha\ra$ be a unit vector.
    	For any CPTP map $\Phi(\rho^{AB})=\sum_iV_i\rho^{AB}V_i^\dag$,
    	there exists a unitary operator $U$
    	acting on $\mH^{ABC}$ such that~\cite{Lindblad1974,Lindblad1975}
    	\be
    	U(A\otimes P_\alpha)U^\dag=\sum_{i,j}V_iAV_j^\dag\otimes|i_c\ra\la j_c|.
    	\ee
    	It is clear that
    	\bea
    	{\rm Tr}_C\left[ I^{AB}\otimes P_iU\left( \rho^{AB}\otimes P_\alpha\right) U^\dag I^{AB}\otimes P_i\right] \quad\quad\nonumber\\
    	\quad\quad=V_i\rho^{AB}V_i^\dag\equiv\Phi_i\left( \rho^{AB}\right) =p_i\rho_i^{AB}.\nonumber
    	\eea
    	According to the proof of Theorem 2 in Ref.~\cite{Vedral1998pra},
    	we only need to verify that
    	if 
    	\bea\label{E_r}
    	\sum_ip_iS\left( \rho_i^{AB}/p_i||\sigma_i^{AB}/q_i\right)
    	=S(\rho^{AB}||\sigma^{AB})
    	\eea
    	holds for any $\rho^{AB}$ and $\sigma^{AB}$, then 
    	\be\label{E_r2}
    	\Phi_i(X)\equiv V_iXV_i^\dag=p_iUXU^\dag
    	\ee for some unitary operator $U$, where $q_i\sigma_i^{AB}=\Phi_i(\sigma^{AB})$.
    	Note that
    	\begin{eqnarray}
        &&\sum_ip_iS( \rho_i^{AB}/p_i||\sigma_i^{AB}/q_i)\nonumber\\
        &\leq&\sum_ip_iS( \rho_i^{AB}/p_i||\sigma_i^{AB}/q_i)+\sum_ip_i\ln\dfrac{p_i}{q_i}\nonumber\\
        &=&\sum_iS( p_i\rho_i^{AB}||q_i\sigma_i^{AB}) \nonumber\\
         &\leq&\sum_iS[ {\rm Tr}_C\{ I^{AB}\otimes P_iU( \rho^{AB}\nonumber\\
         &&\otimes P_\alpha) U^\dag I^{AB}\otimes P_i\} ||{\rm Tr}_C\{  I^{AB}  \nonumber \\
         && \otimes P_iU(  \sigma^{AB}\otimes P_\alpha) U^\dag I^{AB}\otimes P_i\} ] \nonumber\\
    	&\leq&\sum_iS[ I^{AB}\otimes P_iU( \rho^{AB}\otimes P_\alpha) U^\dag I^{AB}\otimes P_i||I^{AB}\nonumber\\
    	&&\otimes P_iU( \sigma^{AB}\otimes P_\alpha) U^\dag I^{AB}\otimes P_i]\nonumber\\
    	&=&S\left[U\left( \rho^{AB}\otimes P_\alpha\right) U^\dag ||U\left( \sigma^{AB}\otimes P_\alpha\right) U^\dag \right]\nonumber\\
    	&=&S\left( \rho^{AB}||\sigma^{AB}\right) ,
    	\end{eqnarray}
        thus~\eqref{E_r} holds and leads to $\sum_ip_i\ln\dfrac{p_i}{q_i}=0$, which is equivalent to 
        $p_i=q_i$ for any $i$.
        Therefore $\Phi_i$ has the form as in~\eqref{E_r2}.
        Taking $V_j=V_j^A\otimes V_j^B$, the proof is completed.
    \end{proof}
		
	The squashed entanglement $E_{\rm sq}$~\cite{Christandl2004jmp}
	is an additive entanglement monotone and has a nice 
	operational meaning.
	For any state $\rho^{AB}\in\mS^{AB}$, $E_{\rm sq}$ is defined by~\cite{Christandl2004jmp}
	\be
	E_{sq}(\rho^{AB})\equiv\inf\limits_{E}\left\lbrace \frac{1}{2}I(A;B|E): {\rm Tr}_E\rho^{ABE}=\rho^{AB}\right\rbrace,
	\ee
	where $I(A;B|E)=S(\rho^{AE})+S(\rho^{BE})-S(\rho^{ABE})-S(\rho^{E})$, $S(\cdot)$ denotes the von Neumann entropy and the infimum is taken over all extensions of 
	$\rho^{ABE}$ of $\rho^{AB}$.
	 We show below that $E_{\rm sq}$
	is also a SEM with the assumption that it can be attained by
	some optimal extension [i.e., $E_{\rm sq}(\rho^{AB})=\frac{1}{2}I(A;B|E)$ for some extension $\rho^{ABE}$].
	Note that, if there does not exist some optimal extension, whether or not $E_{sq}$ is a SEM remains open since
	it is defined in terms of the infimum process over all states extension which cannot give an accurate
	equality between the state and its extension state for the conditional mutual information.
	However, we still do not know such an extension exists or not for any state~\cite{Christandl2004jmp}.
	
	\begin{theorem}\label{7}
		If $E_{\rm sq}(\rho^{AB})$ can be attained by optimal extension for any state $\rho^{AB}\in\mS^{AB}$, then $E_{\rm sq}$ is a SEM.
	\end{theorem}

	\begin{proof}
		From the proof of Proposition 3 in Ref.~\cite{Christandl2004jmp}, if
		$E_{sq}(\rho)=\sum_kp_kE_{sq}(\rho_k')$ and the associated LOCC is stochastic,
		then we must have $I(\tilde{A}';\tilde{B}|\tilde{E})=0$ (we use the same notations as in Ref.~\cite{Christandl2004jmp}), it follows that $\tilde{\rho}^{A'BE}$ is a Markov state according to the structure of states that satisfying the strong subadditivity of entropy~\cite{HaydenJozaPetsWinter}, a contradiction.
		Thus the LOCC is a local unitary operation or a convex mixture of local unitary operations.
	\end{proof}

	At last, we present a list of the properties of all entanglement measures
	that are well-known by now for convenience of readers (see Table~\ref{tab:table1}).
	As one might expect, almost all the entanglement measures are decreasing strictly under LOCC on average for pure states.
	In addition, one can see from the table that, apart from the strict concavity of the associated function $h$, monogamy
	is another property that is also closely related with the strict monotonicity of LOCC.
	We also found that other properties, such as additivity, convexity and faithfulness, seem not to be the nature of 
	the entanglement measures so far.

	To summarize, we
	explored the action of entanglement under LOCC
	for many entanglement measures so far, and we showed
	that the axiomatic definition of entanglement monotone
	can be improved: $E$ is defined to be an entanglement monotone
	if it is convex, vanishes on separable states, and decreases \textit{strictly} on average under LOCC in the sense of~\eqref{strict1}.  
	Together with the result in Ref.~\cite{GG2},
	our results here support the conclusion that 
	entanglement is monogamous.
	But we still can not prove whether the squashed entanglement (it is defined via the infimum over all extensions), entanglement of distillation, and 
	the entanglement cost are strict entanglement monotones or not.

	\begin{acknowledgements}	
		The author is very grateful to the referees for their constructive suggestions.  
		Y.G is supported by the Natural Science Foundation of Shanxi Province
		under Grant No. 201701D121001, the National Natural
		Science Foundation of China under Grant No. 11301312 
		 and the Program for the Outstanding Innovative Teams of Higher Learning Institutions of Shanxi.
	\end{acknowledgements}
	

\end{document}